\documentclass[10pt,b5paper,twoside]{padeu}  %%%% do not change this line
\usepackage{epsfig,natbib_padeu}             %%%% do not change this line
\usepackage{amsmath}

\begin{document}

\title{Weak gravitational lensing in brane-worlds}
\titlerunning{Weak gravitational lensing in brane-worlds}
\author{L\'{a}szl\'{o} \'{A}. Gergely\inst{1}, 
	  Barbara Dar\'{a}zs\inst{2}}
\authorrunning{L. \'{A}. Gergely, B. Dar\'{a}zs}
\institute{Departments of Theoretical and Experimental Physics, University of Szeged,
6720 Szeged, D\'{o}m t\'{e}r 9, Hungary}
\email{\inst{1}gergely@physx.u-szeged.hu, 
       \inst{2}dbarbi@titan.physx.u-szeged.hu}

%________________________________________________________________

\abstract{
We derive the deflection angle of light rays caused by a brane black hole
with mass $m$ and tidal charge $q$ in the weak lensing approach, up to the
second order in perturbation theory. We point out when the newly derived
second order contributions become important.
}

\maketitle

%________________________________________________________________

\section{Introduction}

The possibility of allowing gravitation to exist in a more than
four-dimensional non-compact space-time [\citet{RS}], while keeping the other
interactions locked in four space-time dimensions, has raised interesting
new perspectives in the solvability of the hierarchy problem and in
cosmological evolution. This hypothesis has led to alternative explanations
for both dark energy (see for example [\citet{d2})] and dark matter ([\citet
{MakHarko}], [\citet{Kar}] and [\citet{Pal}]). The simplest so-called brane-world model is
five-dimensional. Gravitational dynamics on the four-dimensional brane is
governed by a modified Einstein equation, derived in full generality in [\citet
{Decomp}].

Gravitational lensing is one of the means by which the existence of
brane-worlds can in principle be tested. A recent review in the topic can be
found in [\citet{MajumdarMukherjee}]. In the context of brane-worlds, both weak 
[\citet{KarSinha}], [\citet{MM}] and strong [\citet{Whisker}] gravitational lensing
were discussed.

Black holes on the brane are described by the \textit{tidal} charged black
holes, derived in [\citet{tidalRN}]:
\begin{equation}
ds^{2}=-f\left( r\right) dt^{2}+f^{-1}\left( r\right) dr^{2}+r^{2}\left(
d\theta ^{2}+\sin ^{2}\theta d\varphi ^{2}\right) \,.  \label{RN}
\end{equation}
The metric function $f$ is given as 
\begin{equation}
f\left( r\right) =1-\frac{2m}{r}+\frac{q}{r^{2}}\,.
\end{equation}
These black holes are characterized by two parameters: their mass $m$ and
their tidal charge $q$. The latter arises from the bulk Weyl curvature (more
exactly, from its ''electric'' part as compared to the brane normal).

Formally the metric (\ref{RN}) is the Reissner-Nordstr\"{o}m solution of a
spherically symmetric Einstein-Maxwell system in general relativity. There,
however the place of the tidal charge $q$ is taken by the square of the 
\textit{electric} charge $Q$. Thus $q=Q^{2}$ is always positive, when the
metric (\ref{RN}) describes the spherically symmetric exterior of an
electrically charged object in general relativity. By contrast, in brane-world theories the metric
(\ref{RN}) allows for any $q$.

The case $q>0$ is in full analogy with the general relativistic
Reissner-Nordstr\"{o}m solution. For $q<m^{2}$ it describes tidal charged
black holes with two horizons at $r_{h}=m\pm \sqrt{\left( m^{2}-q\right) }$,
both below the Schwarzschild radius. For $q=m^{2}$ the two horizons coincide
at $r_{h}=m$ (this is the analogue of the extremal Reissner-Nordstr\"{o}m
black hole). In these cases it is evident that the gravitational deflection
of light and gravitational lensing is decreased by $q$. Finally there is a
new possibility forbidden in general relativity due to physical
considerations on the smallness of the electric charge. This is $q>m^{2}$
for which the metric (\ref{RN}) describes a naked singularity. Such a
situation can arise whenever the mass $m$ of the brane object is of small enough,
compared to the effect of the bulk black hole generating Weyl curvature, and
as such, tidal charge. Due to its nature, the tidal charge$\ q$ should be a
more or less global property of the brane, which can contain many black
holes of mass $m\geq \sqrt{q}$ and several naked singularities with mass $m<
\sqrt{q}$.

For any $q<0$ there is only one horizon, at $r_{h}=m+\sqrt{\left(
m^{2}+\left| q\right| \right) }$. For these black holes, gravity is
increased on the brane by the presence of the tidal charge [\citet{tidalRN}].
Light deflection and gravitational lensing are stronger than for the
Schwarzschild solution.

The metric (\ref{RN}) also describes compact stellar objects. In this case
one does not have to worry about the existence or location of horizons, as
they would lie inside the star, where an interior solution should replace
the metric (\ref{RN}). The generic feature that a positive (negative) tidal
charge is weakening (strengthening) gravitation on the brane, is kept.

In this paper we derive the deflection angle of light rays caused by
brane black holes with tidal charge (\ref{RN}). Generalizing previous
approaches [\citet{KarSinha}], [\citet{MM}], we carry on this computation up to
the second order in the weak lensing parameters. As the metric (\ref{RN}) is
static, we consider only the second order gravielectric contributions, but
no gravimagnetic contributions, which are of the same order and would appear
due to the movement of the brane black holes. Gravimagnetic effects in the
general relativistic approach were considered in [\citet{SchaferBartelmann}]. 

\section{Light propagation}

Light follows null geodesics of the metric (\ref{RN}). Its equations of
motion can be derived either from the geodesic equations, or from the
Lagrangian given by $2\mathcal{L}=\left( ds^{2}/d\lambda ^{2}\right) $ [\citet
{Straumann}] ($\lambda $ being a parameter of the null geodesic curve). Due
to spherical and reflectional symmetry across the equatorial plane, $\theta
=\pi /2$ can be chosen. Thus
\begin{equation}
0=2\mathcal{L}=-f\left( r\right) \dot{t}^{2}+f^{-1}\left( r\right) \dot{r}
^{2}+r^{2}\dot{\varphi}^{2}\,.  \label{Lag}
\end{equation}
(A dot represents derivative with respect to $\lambda$.) 
The cyclic variables $t$ and $\varphi $ lead to the constants of motion $E$
and $L$ 
\begin{equation}
E=f\dot{t}\,,\qquad L=r^{2}\dot{\varphi}\,.  \label{first_integrals}
\end{equation}
By inserting these into Eq. (\ref{Lag}), passing to the new radial variable $
u=1/r$ and introducing $\varphi $ as a dependent variable, we obtain
\begin{equation}
\left( u^{\prime }\right) ^{2}=\frac{E^{2}}{L^{2}}-u^{2}f\,\left( u\right)
\,,  \label{radial}
\end{equation}
where a prime refers to differentiation with respect to $\varphi $. $
\allowbreak $

Unless $u^{\prime }=0$ (representing a circular photon orbit),
differentiation of Eq. (\ref{radial}) gives
\begin{equation}
u^{\prime \prime }=-uf\,-\frac{u^{2}}{2}\frac{df}{du}\,\,,  \label{radial2}
\end{equation}
For $f=1$, when there is no gravitation at all (the metric (\ref{RN})
becomes flat), the above equation simplifies to $u^{\prime \prime }+u=0$,
which is solved for $u=u_{0}=b^{-1}\cos \varphi $. The impact parameter $b$
represents the closest approach of the star on the straight line orbit
obtained by disregarding the gravitational impact of the star (this is the
viewpoint an asymptotic observer will take, as the metric (\ref{RN}) is
asymptotically flat). The polar angle $\varphi $ is measured from the line
pointing from the centre of the star towards the point of closest approach.
With$\ u^{\prime }=0$ at the point of closest approach, given in the
asymptotic limit by $u=b^{-1}$, Eq. (\ref{radial}) with $m=0=q$ gives $b=L/E$
.

\section{Perturbative solution}

Eq. (\ref{radial2}), written in detail, gives
\begin{equation}
u^{\prime \prime }+u=3mu^{2}-2qu^{3}\,\,.  \label{lensing}
\end{equation}
For studying weak lensing, we look for a perturbative solution in series of
the small parameters 
\begin{equation}
\varepsilon =mb^{-1}\,\qquad \text{and}\qquad \eta =qb^{-2}  \label{params}
\end{equation}
in the form
\begin{equation}
u=b^{-1}\cos \varphi +\varepsilon u_{1}+\eta v_{1}+\varepsilon
^{2}u_{2}+\eta ^{2}v_{2}+\varepsilon \eta w_{2}+\mathcal{O}\left(
\varepsilon ^{3},\eta ^{3},\varepsilon \eta ^{2},\varepsilon ^{2}\eta
\right) \,.  \label{ansatz}
\end{equation}
The index on the unknown functions $u_{1},\,u_{2},\,v_{1},\,v_{2}$ and $
w_{2} $ counts the perturbative order in which they appear. By inserting Eq.
(\ref{ansatz}) into the weak lensing equation (\ref{lensing}) we obtain the
relevant differential equations for the unknown functions. Up to the second
order in both small parameters these are: 
\begin{eqnarray}
\varepsilon &:&\qquad u_{1}^{\prime \prime }+u_{1}=3b^{-1}\cos ^{2}\varphi
\,,  \label{ep} \\
\eta &:&\qquad v_{1}^{\prime \prime }+v_{1}=-2b^{-1}\cos ^{3}\varphi \,,
\label{et} \\
\varepsilon ^{2} &:&\qquad u_{2}^{\prime \prime }+u_{2}=3u_{1}\left[
u_{1}\left( m-2qb^{-1}\cos \varphi \right) +2\cos \varphi \right] \,,
\label{ep2} \\
\eta ^{2} &:&\qquad v_{2}^{\prime \prime }+v_{2}=3v_{1}\left[ v_{1}\left(
m-2qb^{-1}\cos \varphi \right) -2\cos ^{2}\varphi \right] \,,  \label{et2} \\
\varepsilon \eta &:&\qquad w_{2}^{\prime \prime }+w_{2}=6\left[
u_{1}v_{1}\left( m-2qb^{-1}\cos \varphi \right) +v_{1}\cos \varphi
-u_{1}\cos ^{2}\varphi \right].  \label{epet}
\end{eqnarray}
$\allowbreak $The first order equations are solved for
\begin{eqnarray}
u_{1} &=&\frac{b^{-1}}{2}\left( 3-\cos 2\varphi \right) \,,  \label{u1} \\
v_{1} &=&-\frac{b^{-1}}{16}\left( 9\cos \varphi -\cos 3\varphi +12\varphi
\sin \varphi \right) \,.  \label{v1}
\end{eqnarray}
Thus, both $mu_{1}$ and $mv_{1}$ are of order $\varepsilon $, while both $
qb^{-1}u_{1}$ and $qb^{-1}v_{1}$ are of order $\eta $. In consequence, all
these terms drop out from Eqs. (\ref{ep2})-(\ref{epet}), which are then
solved for 
\begin{eqnarray}
u_{2} &=&\frac{3b^{-1}}{16}\left( 10\cos \varphi +\cos 3\varphi +20\varphi
\sin \varphi \right) \,,  \label{u2} \\
v_{2} &=&\frac{b^{-1}}{256}\bigl( 192\cos \varphi -48\cos 3\varphi +\cos
5\varphi +384\varphi \sin \varphi \notag\\
&&\qquad\qquad\qquad -36\varphi \sin 3\varphi -72\varphi
^{2}\cos \varphi \bigr) \,,  \label{v2} \\
w_{2} &=&\frac{b^{-1}}{16}\left( -87+40\cos 2\varphi -\cos 4\varphi
+12\varphi \sin 2\varphi \right) \,.  \label{w2}
\end{eqnarray}
With this, we have found the generic solution of Eq. (\ref{lensing}), up to
the second order in both small parameters.

Far away from the lensing object $u=0$ and $\,\varphi =\pi /2+\delta \varphi
/2$, where $\delta \varphi $ represents the angle with which the light ray
is bent by the object with mass $m$ and tidal charge $q$. In our
second-order approach it has the form:\ 
\begin{equation}
\delta \varphi =\varepsilon \alpha _{1}+\eta \beta _{1}+\varepsilon
^{2}\alpha _{2}+\eta ^{2}\beta _{2}+\varepsilon \eta \gamma _{2}+\mathcal{O}
\left( \varepsilon ^{3},\eta ^{3},\varepsilon \eta ^{2},\varepsilon ^{2}\eta
\right) \,.
\end{equation}
A power series expansion of the solution (\ref{ansatz}) then gives the
coefficients of the above expansion, and the deflection angle becomes:
\begin{equation}
\delta \varphi =4\varepsilon -\frac{3\pi }{4}\eta +\frac{15\pi }{4}
\varepsilon ^{2}+\frac{105\pi }{64}\eta ^{2}-16\varepsilon \eta \,.
\label{deflection}
\end{equation}
The first three terms of this expansion were already given in [\citet{Hobill}]
for the Reissner-Nordstr\"{o}m black hole. There, however the argument that $
\eta $ is of $\epsilon ^{2}$ order was advanced. In brane-worlds there is no
a priori reason for considering only small values of the tidal charge, thus
we have computed the deflection angle (\ref{deflection}) containing all
possible contributions up to second order in both parameters.

The deflection angle however is given in terms of the Minkowskian impact
parameter $b$. It would be useful to write this in term of the distance of
minimal approach $r_{\min }$ as well. The minimal approach is found by
inserting the values $u=1/r_{\min }$ and $\varphi =0$ in Eq. (\ref{ansatz}): 
\begin{equation}
r_{\min }=b\left( \allowbreak \allowbreak 1-\varepsilon +\frac{1}{2}\eta -
\frac{17}{16}\varepsilon ^{2}-\frac{81}{256}\eta ^{2}\allowbreak
+2\varepsilon \eta \right) \,.
\end{equation}
Inverting this formula gives to second order accuracy (the small parameters
being now $m/r_{\min }$ and $q/r_{\min }^{2}$):
\begin{equation}
\frac{1}{b}=\frac{1}{r_{\min }}\left( \allowbreak \allowbreak 1-\frac{m}{
r_{\min }}+\allowbreak \allowbreak \frac{q}{2r_{\min }^{2}}-\frac{m^{2}}{
16r_{\min }^{2}}+\frac{47q^{2}}{256r_{\min }^{4}}+\frac{mq}{2r_{\min }^{3}}
\right) \,.
\end{equation}
As the deflection angle consists only of first and second order
contributions, the above formula is needed only to first order for
expressing $\delta \varphi $ in terms of the minimal approach: 
\begin{equation}
\delta \varphi =\allowbreak \allowbreak \frac{4m}{r_{\min }}-\frac{3\pi q}{
4r_{\min }^{2}}\allowbreak \allowbreak +\frac{\left( 15\pi -16\right) m^{2}}{
4r_{\min }^{2}}+\frac{57\pi q^{2}}{64r_{\min }^{4}}+\frac{\left( 3\pi
-28\right) mq}{2r_{\min }^{3}}\,.  \label{deflection2}
\end{equation}
The first three terms again agree with the ones given in [\citet{Hobill}], for $
q=Q^{2}$.

\section{Concluding remarks}

In this paper we have computed the deflection angle caused by a tidal
charged brane black hole /\ naked singularity / star, up to second order in
the two small parameters, related to the mass and tidal charge of the
lensing object.

As already remarked in [\citet{Sereno}], the electric charge of the
Reissner-Nordstr\"{o}m black hole decreases the deflection angle, as compared
to the Schwarz\-schild case. The same is true for a positive tidal charge. In
brane-worlds, however there is no upper limit for $q$ as compared to $m$.
Thus for small mass brane black holes / naked singularities / stars the
condition $16mr_{\min }=3\pi q$ could be obeyed. In this case the first
order contributions to the deflection angle cancel and the three second
order terms of $\delta \varphi $ give the leading effect to the weak lensing.

Furthermore, $16mr_{\min }<3\pi q$ could be obeyed, leading to a \textit{%
negative} deflection angle, at least to first order. That would mean that
rather than magnifying distant light sources, such a lensing object will
demagnify them.

By contrast, a negative tidal charge can considerably increase the lensing
effect. Therefore a negative tidal charge could be responsible at least for
part of the lensing effects attributed at present to dark matter.

\section{Acknowledgement}

This work was supported by OTKA grants no. T046939 and TS044665.

\end{document}